\documentclass{aims}
\usepackage{amsmath}
 \usepackage{booktabs}
  \usepackage{paralist}
  \usepackage{bm}
  \usepackage{graphics} 
  \usepackage{epsfig} 
 \usepackage[colorlinks=true]{hyperref}
\hypersetup{urlcolor=blue, citecolor=red}

\def\currentvolume{X}
 \def\currentissue{X}
  \def\currentyear{200X}
   \def\currentmonth{XX}
    \def\ppages{X--XX}

\theoremstyle{definition}

\title[Integration of variational equations]
      {Comparing the efficiency of numerical techniques for the integration of
variational equations}

\author[Enrico Gerlach and Charalampos Skokos]{}

\subjclass{Primary: 37M25, 65P10; Secondary: 70H07, 37J99, 65P20.}
 \keywords{Variational equations, chaos, SALI method, H\'enon-Heiles system.}

 \email{enrico.gerlach@tu-dresden.de}
 \email{hskokos@pks.mpg.de}

\thanks{The first author is supported by the DFG research unit FOR584--P3.}

\begin{document}
\maketitle

\centerline{\scshape Enrico Gerlach }
\medskip
{\footnotesize
 \centerline{Lohrmann Observatory, Technical University Dresden, D-01062
Dresden, Germany}
} 

\medskip

\centerline{\scshape Charalampos Skokos}
\medskip
{\footnotesize
 \centerline{Max Planck Institute for the Physics of Complex Systems}
   \centerline{N\"othnitzer Str.~38, D-01187 Dresden, Germany}
}

\bigskip

 \centerline{(Communicated by the associate editor name)}

\begin{abstract}
  We present a comparison of different numerical techniques for the
  integration of  variational equations. The methods presented can
  be applied to any autonomous Hamiltonian system whose kinetic energy
  is quadratic in the generalized momenta, and whose potential is a
  function of the generalized positions. We apply the various
  techniques to the well-known H\'enon-Heiles system, and use the
  Smaller Alignment Index (SALI) method of chaos detection to evaluate
  the percentage of its chaotic orbits. The accuracy and the speed of
  the integration schemes in evaluating this percentage are used to
  investigate the numerical efficiency of the various techniques.
\end{abstract}

\section{Introduction}

The determination of the stability of motion is of great importance
when investigating nonlinear dynamical systems. To distinguish
correctly between regular and chaotic motion, several different methods
have been developed during the years.  Most of these techniques, such as
the maximal Lyapunov exponent \cite{S10}, the fast Lyapunov indicator
\cite{FLG1997} or the Smaller Alignment Index (SALI) \cite{S01}, rely
on the study of the time evolution of deviation vectors from a given
orbit to discriminate between the two regimes. The time evolution of
these vectors is governed by the so-called variational equations.

Besides the correct determination of the regular or chaotic nature of
individual orbits, in many cases, statistical statements over a large
region of the phase space are also needed. For example, in order to
determine the percentage of regular and chaotic orbits in a given
system, the characterization of many orbits is required.  In addition
to the accurate computation of chaos indicators also for such tasks
the CPU time needed to perform these computations becomes very
important. In this work we compare different numerical techniques for
the integration of the variational equations, concentrating on their
accuracy and computational speed.

The paper is organized as follows: in section \ref{sec:2} we describe
the general layout of our investigation. We concentrate our study on a
simple two degrees of freedom Hamiltonian system: the well-known
H\'enon-Heiles system \cite{HH64}, which is presented in section
\ref{sec:HH}. In our study we use the SALI method of chaos detection,
which is presented in section \ref{sec:sali}. To solve the equations
of motion of the H\'enon-Heiles system and the associated variational
equations one has to employ numerical methods. Any non-symplectic
general-purpose integrator can be used for this task.  In sections
\ref{sec:DOP853_t} and \ref{sec:taylor_m} we present two such
techniques, which we use in our study. In \cite{paper1} it was shown
that it is also possible to use methods based on symplectic
integration techniques to solve these equations. In section
\ref{sec:tm} we describe shortly the most efficient of these
techniques: the so-called \emph{tangent map (TM) method}.  A numerical
procedure to obtain relatively fast information on the nature of
orbits for a large set of initial conditions is then given in section
\ref{sec:pss}. In section \ref{sec:results} we present our numerical
results for individual orbits, as well as global results for the whole
system. The summary and the conclusions of our study are found in
section \ref{sec:conclusions}.

\section{\label{sec:2}Numerical integration of variational equations}
\subsection{\label{sec:HH}H\'enon-Heiles system}

The Hamiltonian function of the  H\'enon-Heiles system \cite{HH64} is
\begin{equation}
H(q_1,q_2,p_1,p_2) = \frac{1}{2} (p_1^2+p_2^2) + \frac{1}{2} (q_1^2+q_2^2) +
q_1^2 q_2 -
\frac{1}{3} q_2^3,
\label{eq:HH}
\end{equation}
with $q_1$, $q_2$ being the generalized coordinates and $p_1$, $p_2$
the conjugate momenta.  The orbit evolution is given by the standard
Hamilton equations of motion
\begin{equation}
 \dot q_i = \frac{\partial H}{\partial p_i}\qquad\mathrm{and}\qquad \dot p_i =
-\frac{\partial H}{\partial q_i},\qquad i=1,2,
\label{eq:HHeq}
\end{equation}
where the dot denotes derivation with respect to time $t$. The time
evolution of the variations $\delta q_i,\delta p_i$ (which can be
considered as coordinates of a deviation vector) is governed by the
variational equations, given by
\begin{equation}
 \dot{\delta q_i} = \delta p_i\qquad\mathrm{and}\qquad \dot{\delta p_i} = -\sum
_{j=1}^2\frac{\partial^2 H}{\partial q_i\partial q_j}\delta q_j,\qquad i=1,2.
\label{eq:HHvareq}
\end{equation}
Eqs.~(\ref{eq:HHeq}) and (\ref{eq:HHvareq}) form a coupled system of
ordinary differential equations. It should be noted that the solution
of (\ref{eq:HHvareq}) depends explicitly on the solution of
(\ref{eq:HHeq}), i.~e.~on the reference orbit $q_i(t),p_i(t)$, and
thus Eq.~(\ref{eq:HHvareq}) cannot be solved independently from
Eq.~(\ref{eq:HHeq}).

\subsection{\label{sec:sali}SALI method}

The evaluation of the SALI is an efficient and simple method to
determine the regular or chaotic nature of orbits in dynamical
systems. The SALI was proposed in \cite{S01} has since been
successfully applied in order to distinguish between regular and
chaotic motion both in symplectic maps and Hamiltonian flows
\cite{SABV03,SABV04,SESS04,PBS04,BS05,MSAB08,SHC09}. For the
computation of the SALI of a given orbit, one has to follow the time
evolution of the orbit itself and also of two deviation vectors
$V_{1}(t),V_{2}(t)$, which initially point in two different
directions. Then, according to \cite{S01} the SALI is defined as
\begin{equation}\label{eq:SALI:2}
SALI(t)=min
\left\{\left\|\hat{V}_{1}(t)+\hat{V}_{2}(t)\right\|,\left\|\hat{V}_{1}(t)-\hat{V
}_{2}(t)\right\|\right\}, 
\end{equation}
where $\| \cdot \|$ denotes the usual Euclidean norm and
$\hat{V}_{i}$, $i=1,2$ are normalized vectors with norm equal to 1.

The SALI has a completely different behavior for regular and chaotic
orbits, and this allows us to clearly distinguish between them. In
particular, the SALI fluctuates around a non-zero value for regular
orbits, while it tends exponentially to zero for chaotic orbits
\cite{S01,SABV03}, following a rate which depends on the difference of
the two largest Lyapunov exponents \cite{SABV04}.

\subsection{\label{sec:num_methods}Used numerical methods}
\subsubsection{\label{sec:DOP853_t}DOP853}

The DOP853 \footnote{Freely available from
  \texttt{http://www.unige.ch/\textasciitilde hairer/software.html}.}
integration method belongs to the big class of explicit Runge-Kutta
methods.  This non-symplectic scheme of order 8 is based on the method
of Dormand and Price (see \cite[Sect.~II.5]{Hairer_etal_93}). We use
this integrator to solve the set of differential equations composed of
Eqs.~(\ref{eq:HHeq}) and (\ref{eq:HHvareq}).  Two free parameters,
$\tau$ and $\delta$, are used to control its numerical
performance. The first one defines the time span between two
successive outputs of the computed solution. After each step of length
$\tau$ the deviation vectors are renormalized and the value of SALI is
computed.  For the duration of each step $\tau$, the integrator
adjusts its own internal time step in order to keep the local one-step
error smaller than a user-defined threshold value $\delta$. For the
DOP853 integrator the estimation of this local error and the step size
control is based on embedded formulas of orders 5 and 3.

\subsubsection{\label{sec:taylor_m}Taylor methods}

The basic idea of the so-called Taylor series methods (for details see
for example \cite[Sect.~I.8]{Hairer_etal_93} and references therein)
is to approximate the solution at time $t_i+\tau$ of a given
$s$-dimensional initial value problem
\begin{equation}
 \frac{\mathrm d \bm y(t)}{\mathrm d t} = \bm f(\bm y(t))\qquad\bm y\in \mathbb
R^s,\, t\in \mathbb R
\end{equation}
from the $n$th degree Taylor series of $\bm y(t)$ at $t=t_i$ as
\begin{equation}
 \bm y(t_i+\tau)\simeq \bm y(t_i)+ \tau\frac{\mathrm d\bm y(t_i)}{\mathrm
dt}+\frac{\tau^2}{2!}\frac{\mathrm d^2\bm y(t_i)}{\mathrm dt^2}+\ldots+
\frac{\tau^n}{n!}\frac{\mathrm d^n\bm y(t_i)}{\mathrm dt^n}.
\end{equation}
The computation of the derivatives is commonly done using automatic
differentiation (see for example \cite{jorba}).

In our study we use two different public available implementations of
the Taylor method: \emph{TIDES} \footnote{Freely available from
  \texttt{http://gme.unizar.es/software/tides}.}  \cite{barrio} and
\emph{TAYLOR} \footnote{Freely available from
  \texttt{http://www.maia.ub.es/\textasciitilde
    angel/taylor/software/}.}  \cite{jorba}. Both methods have
internal automatic order and step size computation to ensure the
user-defined local one-step error $\delta$. Also here the parameter
$\tau$ defines the step size, after which the renormalization of the
deviation vectors and the computation of SALI is done.

The whole testbed of our work is written using the FORTRAN programming
language exploiting extended double precision\footnote{Corresponding
  to 18 significant digits or equivalently to a machine accuracy of
  $\approx 10^{-19}$.}. While TIDES offers directly a FORTRAN
integration routine, a wrapper to include the routine written in C had
to be used for TAYLOR. Therefore for the latter only 16 significant
digits were available for the integration.

\subsubsection{\label{sec:tm}TM method}

Besides general-purpose integrators, it is also possible to use
techniques based on symplectic methods to integrate the Hamilton
equations of motion and the corresponding variational equations. This
was shown in \cite{paper1}, where a thorough discussion of possible
methods can be found. The most effective of these techniques, the TM
method, is used in this work.  Let us outline its basic idea, which is
founded on a general result stated for example in \cite{LR01}:
Symplectic integrators can be applied to first order differential
systems $\dot X=LX$, that can be written in the form $\dot
X=(L_A+L_B)X$, where $L,L_A,L_B$ are differential operators defined as
$L_\chi f=\{\chi,f\}$ and for which the two systems $\dot X=L_A X$ and
$\dot X=L_B X$ are integrable. Here $\{f,g\}$ are Poisson brackets of
functions $f(\bm q,\bm p)$, $g(\bm q,\bm p)$ defined as:
\begin{equation}
\{ f,g\}=\sum_{l=1}^{N} \left( \frac{\partial f}{\partial p_l}
\frac{\partial g}{\partial q_l} - \frac{\partial f}{\partial q_l}
\frac{\partial g}{\partial p_l}\right). \label{eq:Poisson}
\end{equation}
The set of Eqs.~(\ref{eq:HHeq}) and (\ref{eq:HHvareq}) is one example
of such a system, because Hamiltonian (\ref{eq:HH}) can be divided
into two integrable parts $A$ and $B$ with $H=A(\bm p)+B(\bm q)$. A
symplectic integrator splits the equations of motion (\ref{eq:HHeq})
into several parts, applying either the operator $L_A$ or $L_B$. These
are the equations of motion of the Hamiltonians $A$ and $B$, which can
be solved analytically, giving explicit mappings over the time step
$c_i\tau$, where the constants $c_i$ are chosen to optimize the
accuracy of the integrator. These mappings can then be combined to
approximate the solution after time step $\tau$. In \cite{paper1} it
was shown that the derivative of these mappings - with respect to the
coordinates and momenta of the system (the so-called tangent maps) - can
be used for the time evolution of deviation vectors or, in other
words, for solving the variational equations (\ref{eq:HHvareq}). We
note that the TM method is called the `global symplectic integrator
method' in \cite{LHC10}.

In \cite{LR01} a family of symplectic integrators called SABA$_n$ and
SBAB$_n$ was introduced, with $n$ being the number of applications of
operators $L_A$ and $L_B$.  These integrators have only positive
intermediate steps and can be used with an additional corrector step
$C$ at the beginning and the end of each step $\tau$ to increase their
accuracy. An integrator of order 4 of this family, namely the
SBAB$_2$C integrator which includes corrector steps, is used in our
investigation. A detailed description of the application of the
SBAB$_2$C integrator for the TM method to the H\'enon-Heiles system 
can be found in \cite{paper1}.

\subsection{\label{sec:pss}Fast PSS method}

Besides information on the chaotic or regular character of individual
orbits, a more global description of dynamical systems is also of
interest. For example, such a study could include the computation of
the percentage of regular/chaotic orbits for a given set of initial
conditions (ICs).  This information requires the integration of the
equations of motion/variational equations, and the computation of
a chaos indicator for the whole set of ICs, which can become a very
hard computational task. In order to address this problem, we implement
a method proposed in \cite{AMS04}, which exploits the Poincar\'{e}
surface of section (PSS) of the system in order to speed up this
computation.

\begin{figure}[htp]
\begin{center}
  \includegraphics[width=0.6\textwidth]{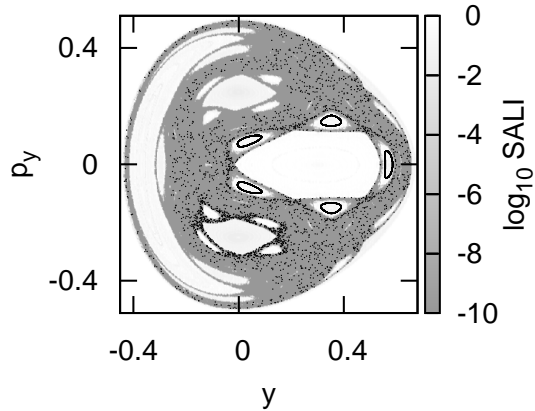}\\
  \caption{The PSS of the H\'enon-Heiles system (\ref{eq:HH}) for
    $q_1=0$ and $p_1\ge 0$.  A grid of $400\times400$ initial
    conditions is used and each initial condition is colored in
    grey-scale according to its SALI value at $t=3000$. Black dots
    forming five closed curves correspond to the regular orbit R1 of
    Fig.~\ref{fig:init5}, while the scattered black dots are
    intersection points of the chaotic orbit C1 of
    Fig.~\ref{fig:init4} with the PSS.}\label{fig:pss}
  \end{center}
\end{figure}

For a fixed value of Hamiltonian (\ref{eq:HH}) (throughout our study
we use always $H=0.125$) we define the PSS of the system as the plane
given by $q_1=0$ and $p_1\ge 0$. Each point in this plane defines a
set of values ($q_2,p_2$). To evaluate the percentage of regular
orbits, one normally computes for each point the value of some chaos
indicator using a dense set of points on the PSS as ICs.  In
Fig.~\ref{fig:pss} we consider a grid of $400\times400$ ICs and color
each one according to its SALI value at $t=3000$.

Each orbit starting from any IC intersects the PSS in many points, and
so, its SALI value can be attributed to all orbits having these
intersection points as ICs. Therefore all these ICs do not have to be
integrated separately. This procedure decreases drastically the CPU
time needed for the global description of the system's chaoticity. We
refer to this approach as the \emph{fast PSS method}.

\section{\label{sec:results}Numerical results}
\subsection{Individual orbits}

Let us first investigate, how well the different methods described in
section \ref{sec:num_methods} can determine the nature of individual
orbits. We use these methods to integrate Eqs.~(\ref{eq:HHeq}) and
(\ref{eq:HHvareq}), and then we compute the evolution of the SALI in
order to determine the nature of the orbit. Unless otherwise stated, we
always renormalize the deviation vectors after each time step of
length $\tau=0.05$. For the non-symplectic routines we adopt a
one-step accuracy of $\delta=10^{-5}$.

\begin{figure}[htb]
\begin{center}
  \includegraphics[width=\textwidth]{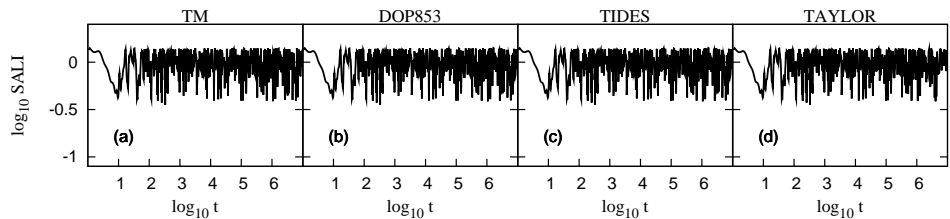}\\
  \caption{Time evolution of the SALI for the regular orbit R1 with
    initial conditions $q_1=0$, $p_1\approx0.2334$, $q_2=0.558$,
    $p_2=0$.}\label{fig:init5}
  \end{center}
\end{figure}

\begin{figure}[htb]
\begin{center}
  \includegraphics[width=\textwidth]{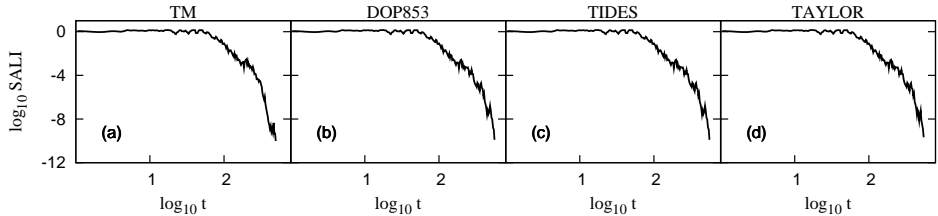}\\
  \caption{Time evolution of the SALI for the chaotic orbit C1 with
    initial conditions $q_1=0$, $p_1\approx0.4208$, $q_2=-0.25$,
    $p_2=0$.}\label{fig:init4}
  \end{center}
\end{figure}

\begin{figure}[htb]
\begin{center}
  \includegraphics[width=\textwidth]{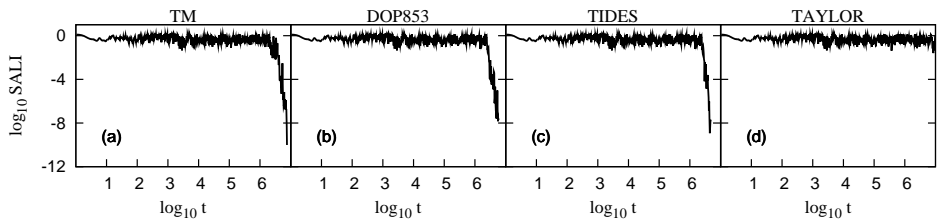}\\
  \caption{Time evolution of the SALI for the sticky chaotic orbit C2
    with initial conditions $q_1=0$, $p_1\approx0.11879$,
    $q_2=0.335036$ and $p_2=-0.385631$.}\label{fig:init1}
  \end{center}
\end{figure}

As representative examples we consider 3 orbits of the H\'enon-Heiles
system with different dynamical behaviors. The evolution of the SALI
for a regular orbit (R1) is presented in Fig.~\ref{fig:init5}, while
in Fig.~\ref{fig:init4} we have similar results for a chaotic orbit
(C1). Finally, in Fig.~\ref{fig:init1} the SALI of a sticky chaotic
orbit (C2) is shown.

\begin{figure}[htp]
\begin{center}
  \includegraphics[scale=.87]{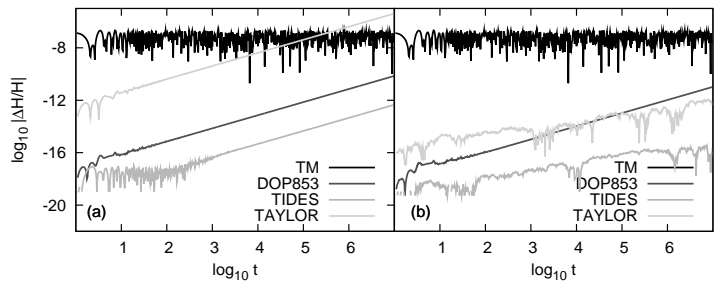}\\
  \caption{The time evolution of the relative energy error $|\Delta
    H/H|$ for the different integration schemes in the case of the
    regular orbit R1. The step size $\tau$ was 0.05 for all
    methods. For all non-symplectic methods we set (a)
    $\delta=10^{-5}$ and (b) $\delta=10^{-16}$. The relative energy
    error for the TM method is the same for both panels and is
    reported only for reference, since for this method no one-step
    accuracy $\delta$ is defined.}\label{fig:energy}
  \end{center}
\end{figure}

From Fig.~\ref{fig:init5} we see that the results obtained for orbit
R1 by the different integration methods are nearly identical. As
theory predicts a constant SALI for regular motion, such behavior is
correctly identified for orbit R1. Information concerning the
numerical performance of various techniques for orbit R1 is given in
Table~\ref{tab:cpu} (throughout this paper the reported CPU times
refer to an Intel Xeon X5660, 2.80~GHz computer).  From the results
reported in this table, we see that differences between the applied
methods appear in their energy conservation properties, as is
indicated by the relative energy error $|\Delta H/H|$, shown in
Fig.~\ref{fig:energy}. For the symplectic algorithm (TM method)
$|\Delta H/H|$ shows fluctuations around $10^{-8}$, while it grows
with time for the other methods as expected (see for example
\cite[Sect.~IX.8]{Hairer_etal_02}). The TAYLOR method has the worst
performance since it is able to conserve the energy only up to an
error level of $\approx 10^{-6}$ for $\delta =10^{-5}$. This is
probably due to the 2 digits less in accuracy that are available for
this method (see section \ref{sec:num_methods}). The best method with
respect to the energy conservation is the TIDES algorithm for which
the relative error is $\approx 10^{-13}$ ($\delta =10^{-5}$) and
$\approx 10^{-16}$ ($\delta =10^{-16}$) at $t=10^7$.  The price paid
for the excellent accuracy of the algorithm is that TIDES requires, in
general, the largest CPU times and the highest orders among the tested
methods.

\begin{table} [!ht]
  \caption{Information on the performance  of the different numerical 
    methods used for the computation of the evolution of the regular orbit R1, of two deviation
    vectors from it and  of its SALI. The step size $\tau$ was always 0.05. The
    order used by the TIDES and TAYLOR methods is determined by these routines for
    each step $\tau$ and is constant for the whole integration.}\label{tab:cpu}
\centering
\footnotesize
\begin{tabular}{rrrcr}
integrator & method & CPU time & Relative energy error &
order \\
\toprule
SBAB$_2$C & TM                  & 04m 02s   & $2\times10^{-8}$  & 4\\
DOP853    & $\delta=10^{-5}$  & 09m 05s & $7\times10^{-11}$ & 8\\
DOP853    & $\delta=10^{-16}$ & 15m 58s & $1\times10^{-11}$ & 8\\
TIDES     & $\delta=10^{-5}$  & 15m 45s & $4\times10^{-13}$ & 10\\
TIDES     & $\delta=10^{-16}$ & 39m 39s & $1\times10^{-16}$ & 23\\
TAYLOR    & $\delta=10^{-5}$  & 15m 00s & $4\times10^{-6}$  & 7\\
TAYLOR    & $\delta=10^{-16}$ & 67m 01s & $4\times10^{-13}$ & 20\\
\bottomrule
\end{tabular}
\end{table}

Orbit C1 is also correctly identified as chaotic by all methods in
less than 1000 time units, within which SALI goes to zero
(Fig.~\ref{fig:init4}). A difference is found in the results for the
sticky chaotic orbit C2 (Fig.~\ref{fig:init1}). Up to $t \approx 10^6$
all methods indicate a regular behavior. It is only afterwards that
SALI goes to zero for the TM, the DOP853 and the TIDES methods,
correctly identifying C2 as a sticky chaotic orbit. For the used
values of $\delta$ and $\tau$ the TAYLOR method does not succeed to
show the decrease of SALI to zero, at least up to $t=10^7$.

\subsection{Global results}

\begin{figure}[htp]
\begin{center}
  \includegraphics[width=0.45\textwidth]{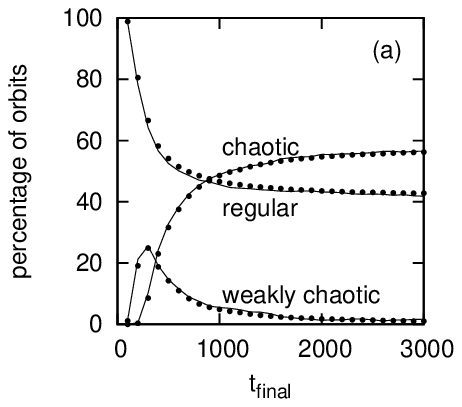} 
 \includegraphics[width=0.45\textwidth]{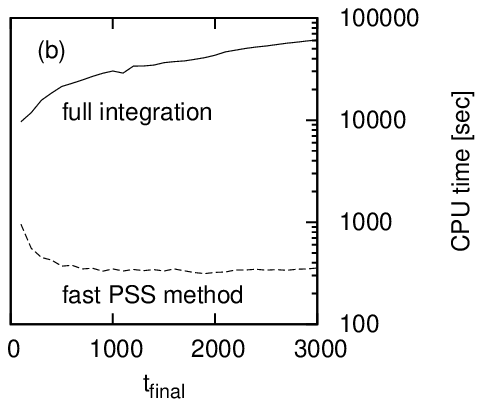}\\
 \caption{(a) Percentage of regular, chaotic and sticky chaotic orbits
   as a function of $t_\mathrm{final}$, when each initial condition is
   integrated until time $t_\mathrm{final}$ (black dots) and when the
   fast PSS method was used (solid lines). The orbits are
   characterized according to their SALI value at time
   $t_\mathrm{final}$ as regular ($\mathrm{SALI}\geq 10^{-4}$),
   chaotic ($10^{-8}>\mathrm{SALI}$) and sticky chaotic
   ($10^{-4}>\mathrm{SALI} \geq 10^{-8}$). A grid of $400\times400$
   initial conditions on the PSS of Fig.~\ref{fig:pss} was used.  The
   integrations of the orbit and the deviation vectors were done by
   the TM method using the SBAB$_2$C integrator with a step size
   $\tau=0.05$. (b) The CPU time needed for the computation of the
   results shown in (a).}\label{fig:compPSS}
  \end{center}
\end{figure}

In order to find the percentage of regular and chaotic orbits of the
H\'enon-Heiles system (\ref{eq:HH}), we compute for a grid of
$400\times400$ ICs on the PSS the SALI value for different final times
$t_\mathrm{final}$.  One could argue that due to the finite resolution
with which the grid of ICs is taken on the PSS, the fast PSS method
(Sec.~\ref{sec:pss}) would not be as accurate as the individual
computation of SALI for each IC. The percentages (over the total
number of ICs compatible with Hamiltonian (\ref{eq:HH})) of regular ($\mathrm{SALI}\geq 10^{-4}$), chaotic
($10^{-8}>\mathrm{SALI}$), and sticky chaotic orbits
($10^{-4}>\mathrm{SALI} \geq 10^{-8}$) obtained by both approaches
using the TM method, are shown in Fig.~\ref{fig:compPSS}(a), while the
required CPU times are reported in Fig.~\ref{fig:compPSS}(b). From the
results of Fig.~\ref{fig:compPSS} we see that both approaches obtain
practically the same values, while the CPU time needed by the fast PSS
method remains considerably smaller with respect to the full
integration of individual orbits. For this reason we apply the fast
PSS method for computing the percentages of regular, chaotic and
sticky chaotic orbits for different values of the time step $\tau$ and
the final time $t_\mathrm{final}$. The obtained results can be found
in Table~\ref{tab:2}.

\begin{table} [!ht]
  \caption{Percentage of  regular, chaotic and sticky chaotic orbits for different
    values of the final time $t_\mathrm{final}$ and the step length $\tau$, for the 
    integration schemes presented in section \ref{sec:num_methods}. A grid of
    $400\times400$ initial conditions on the PSS of Fig.~\ref{fig:pss} is used. 
    $\delta$ was set to $10^{-5}$ for the non-symplectic methods. The required CPU times are reported in the last column.}\label{tab:2}
\centering
\footnotesize
\begin{tabular}{ccrcccr}
$\tau$ & $t_\mathrm{final}$  & method & \% regular & \% sticky chaotic & \%
chaotic & CPU time\\
\toprule
0.50 & $10^4$ & TM  &  40.89 & 0.38 & 58.73  &  0 h  02 min\\
	&  & DOP853 &  40.18 & 0.79 & 59.02  &  0 h  04 min\\
	&  & TIDES  &  40.18 & 0.79 & 59.02  &  0 h  04 min\\
\vspace{1mm}
 	&  & TAYLOR &  40.18 & 0.79 & 59.02  &  0 h  05 min\\
0.50 & $10^5$ & TM  &  40.05 & 0.07 & 59.88  &  0 h  09 min\\
 	&  & DOP853 &  39.71 & 0.37 & 59.92  &  0 h  14 min\\
	 &  & TIDES &  39.22 & 0.49 & 60.29  &  0 h  17 min\\
\vspace{1mm}
 	&  & TAYLOR &  40.19 & 0.47 & 59.34  &  0 h  15 min\\
0.50 & $10^6$ & TM  &  40.00 & 0.00 & 60.00  &  1 h  21 min\\
 	&  & DOP853 &  36.94 & 2.59 & 60.47  &  1 h  29 min\\
	 &  & TIDES &  35.78 & 3.71 & 60.51  &  2 h  06 min\\
\vspace{1mm}
 	&  & TAYLOR &  34.01 & 6.23 & 59.76  &  1 h  15 min\\
0.01 & $10^4$ & TM  &  40.38 & 0.72 & 58.90  &  0 h  19 min\\
 	&  & DOP853 &  40.22 & 0.70 & 59.08  &  0 h  38 min\\
	&  & TIDES  &  40.22 & 0.70 & 59.08  &  1 h  05 min\\
\vspace{1mm}
	&  & TAYLOR &  40.23 & 0.84 & 58.93  &  0 h  59 min\\
0.01 & $10^5$ & TM  &  39.39 & 0.57 & 60.04  &  2 h  02 min\\
 	&  & DOP853 &  39.84 & 0.22 & 59.95  &  4 h  12 min\\
	&  & TIDES  &  39.84 & 0.22 & 59.95  &  7 h  20 min\\
\vspace{1mm}
	&  & TAYLOR &  39.83 & 0.28 & 59.89  &   6 h  43 min\\
0.01 & $10^6$ & TM  &  40.01 & 0.00 & 59.99  &  19 h  26 min\\
 	&  & DOP853 &  40.02 & 0.28 & 59.70  &  39 h  42 min\\
	&  & TIDES  &  40.02 & 0.28 & 59.70  &  68 h  08 min\\
	&  & TAYLOR &  39.95 & 0.21 & 59.84  &  62 h  32 min\\
\bottomrule
\end{tabular}
\end{table}

\begin{figure}[htb]
\begin{center}
  \includegraphics[width=\textwidth]{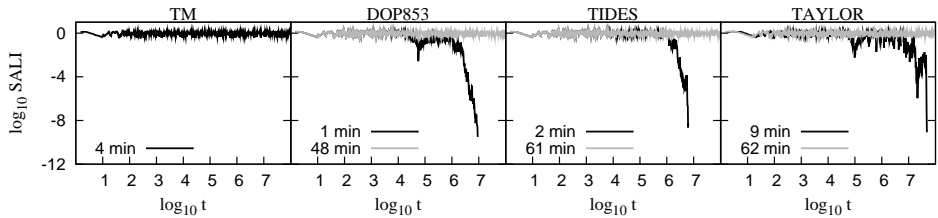}\\
  \caption{Time evolution of the SALI for the regular orbit R1 for
    $\tau=0.50$.  For the DOP853, the TIDES and the TAYLOR
    non-symplectic methods we set $\delta=10^{-5}$ (black line) and
    $\delta=10^{-14}$ (grey line). The required CPU times are given as
    labels of the various lines.}\label{fig:new}
  \end{center}
\end{figure}

From the results of Table~\ref{tab:2} we see that for large values of
$\tau$ and $t_\mathrm{final}$ ($\tau=0.50$ and
$t_\mathrm{final}=10^6$) the non-symplectic methods find $3-4\%$ less
regular orbits than the TM method. In order to understand this
discrepancy we computed for orbit R1 the evolution of the SALI by the
DOP853, the TIDES and the TAYLOR methods with $\tau=0.50$ and
$\delta=10^{-5}$ (Fig.~\ref{fig:new}). From Fig.~\ref{fig:new} we see
that all non-symplectic methods fail to detect correctly the regular
character of the orbit because SALI drops to zero after $t=10^6$. Such
behaviors lead to the increase of the percentage of sticky chaotic
orbits in Table~\ref{tab:2}, since some regular orbits are wrongly
characterized as sticky or chaotic.  Decreasing $\delta$ to values
$\leq 10^{-14}$ solves the problem, as it leads to a correct
identification of the orbit's nature, but also increases the required
CPU time (see Fig.~\ref{fig:new}).

For smaller values of $\tau$ the results obtained from different
techniques are more consistent. For large integration times all
methods give a similar percentage of regular orbits of $\approx
40\%$. TM, DOP853 and TIDES agree here within $0.01\%$. For
$\tau=0.01$ and $t=10^6$ the TM method clearly discriminates between
regular and chaotic orbits, while all other methods still find
$\approx 0.3\%$ of sticky chaotic orbits. But since the SALI of those
orbits will eventually go to zero as well, it can be expected that
also the non-symplectic methods will give the same results as the TM
method when the integration time is increased.

\section{\label{sec:conclusions}Summary and conclusions}

We considered the problem of  fast and accurate integration of the
variational equations of a conservative Hamiltonian system. We
compared different numerical techniques for this task, and applied
them to the H\'enon-Heiles system. We considered non-symplectic
methods of high accuracy; particularly the DOP853 scheme, as well
as the TIDES and TAYLOR packages, which are based on Taylor expansion
techniques. We also applied the TM method, which exploits symplectic
integrators, using in particular the SBAB$_2$C integrator.

The variational equations govern the evolution of small deviations from
a given orbit.  Using the SALI chaos indicator, which is defined
through the time evolution of deviation vectors, we determined the
chaotic or regular nature of individual orbits. In addition, applying
an efficient numerical approach, the so-called fast PSS method, we
were able to rapidly identify regions of order and chaos in the phase
space of the system.

Our numerical results show that the TM method had the best numerical
performance both in accuracy and in speed, especially for large
integration steps, when the other non-symplectic schemes failed to
compute accurately the fraction of regular and chaotic motion. For
moderate integration steps all applied methods gave practically the
same results, with the TM method being always faster.

Among the non-symplectic algorithms the TIDES was the most accurate
one producing similar results as the DOP853 integrator. In many
cases the results of the TAYLOR method were found to be less accurate
than the ones obtained by the other methods, probably due to some
implementation peculiarities of the algorithm.


\medskip
Received xxxx 20xx; revised xxxx 20xx.
\medskip


\begin{thebibliography}{99}

\bibitem{AMS04} 
	\newblock Ch.~Antonopoulos, A.~Manos and Ch.~Skokos,
	\newblock \emph{SALI: an efficient indicator of chaos with application
to 2 and 3 degrees of freedom Hamiltonian systems},
	\newblock  in ``Proc.of the 1st International Conference: From
Scientific Computing to Computational Engineering" (ed. D.T.~Tsahalis), Patras
Univ. Press, (2005), 1292--1298.

\bibitem{barrio} (MR2121808)
	\newblock R.~Barrio,
	\newblock \emph{Performance of the Taylor series method for ODEs/DAEs}
	\newblock Appl.~Math.~Comput., \textbf{163} (2005), 525--545.

\bibitem{BS05} 
\newblock T.~C.~Bountis and Ch.~Skokos,
\newblock \emph{Application of the SALI chaos detection method to accelerator
mappings},
\newblock  Nucl.~Inst.~Meth.~Phys.~Res.~A, \textbf{561} (2006), 173--179.


\bibitem{FLG1997} (MR1472597)
	\newblock C.~Froeschl{\'e}, and E.~Lega and R.~{Gonczi},
	\newblock \emph{Fast Lyapunov Indicators. Application to Asteroidal
Motion}
	\newblock Cel.~Mech.~Dyn.~Astron., \textbf{67} (1997), 41--62.

\bibitem{Hairer_etal_93} (MR1227985)
	\newblock E.~Hairer, S.~P.~N{\o}rsett and G.~Wanner,
	\newblock ``Solving Ordinary Differential Equations. Nonstiff Problems",
	\newblock 2$^{nd}$ edition, Springer Series in Comput.~Math., 1993.

\bibitem{Hairer_etal_02} (MR2221614)
	\newblock E.~Hairer, C.~Lubich and G.~Wanner,
	\newblock ``Geometric Numerical Integration. Structure-Preserving
Algorithms for Ordinary Differential Equations",
	\newblock Springer Series in Comput.~Math., 2002.

\bibitem{HH64} (MR0158746)
\newblock M.~H\'{e}non and C.~Heiles, 
\newblock \emph{The applicability of the third integral of motion: some
numerical experiments}
\newblock Astron.~J.,  \textbf{69} (1964), 73--79.


\bibitem{jorba} (MR2146523)
	\newblock A.~Jorba and M.~Zou,
	\newblock \emph{A Software Package for the Numerical Integration of ODEs
by Means of High-Order Taylor Methods},
	\newblock Experimental Mathematics, \textbf{14} (2005), 99--117.

\bibitem{LR01} (MR1864262)
	\newblock J.~Laskar and P.~Robutel,
	\newblock \emph{High order symplectic integrators for perturbed
Hamiltonian systems},
	\newblock Cel.~Mech.~Dyn.~Astr., \textbf{80} (2001), 39--62.

\bibitem{LHC10} 
	\newblock A.-S.~Libert, C.~Habaux and T.~Carletti,
	\newblock \emph{Symplectic integration of deviation vectors and chaos
determination. Application
to the H\'{e}non-Heiles model and to the restricted three-body problem.},
\newblock MNRAS, \textbf{414} (2011), 659--667.


\bibitem{MSAB08} (MR2479522)
\newblock T.~Manos, Ch.~Skokos, E.~Athanassoula  and T.~Bountis,
\newblock \emph{Studying the global dynamics of conservative dynamical systems
using the SALI chaos detection method}
\newblock Nonlin.~Phenom.~Complex Syst., \textbf{11} (2008), 171--176.


\bibitem{PBS04} 
\newblock P.~Panagopoulos, T.~C.~Bountis and Ch.~Skokos,
\newblock \emph{Existence and stability of localized oscillations in
1-dimensional lattices with soft spring and hard spring potentials}
\newblock  J. Vib. \& Acoust.,  \textbf{126} (2004), 520--527.

\bibitem{SESS04} 
      \newblock A.~Sz\'{e}ll, B.~\'{E}rdi, Zs.~S\'{a}ndor  and B.~Steves,
\newblock \emph{Chaotic and stable behavior in the Caledonian Symmetric
Four-Body problem},
\newblock  MNRAS, \textbf{347} (2004), 380--388.

\bibitem{S01} (MR1871759)
     \newblock Ch.~Skokos,
     \newblock \emph{Alignment indices: A new, simple method for determining the
ordered or chaotic nature of orbits}, 
     \newblock J.~Phys.~A, \textbf{34} (2001), 10029--10043.

\bibitem{S10} 
     \newblock Ch.~Skokos, 
     \newblock \emph{The Lyapunov Characteristic Exponents and their
computation},
     \newblock in Lect.~Notes Phys., \textbf{790} (2010), 63--135.

\bibitem{paper1}
\newblock Ch.~Skokos and E.~Gerlach,
\newblock \emph{Numerical integration of variational equations},
\newblock Phys. Rev. E, \textbf{82} (2010), 036704.

\bibitem{SABV03} 
     \newblock Ch.~Skokos, Ch.~Antonopoulos, T.C.~Bountis and M.N.~Vrahatis,
     \newblock \emph{How does the Smaller Alignment Index (SALI) distinguish
order from chaos?},
     \newblock Prog. Theor. Phys. Supp., \textbf{150} (2003), 439-443.

\bibitem{SABV04} (MR2073606)
     \newblock Ch.~Skokos, Ch.~Antonopoulos, T.C.~Bountis and M.N.~Vrahatis,
     \newblock \emph{Detecting order and chaos in Hamiltonian systems by the
SALI method},
     \newblock J.~Phys.~A, \textbf{37} (2004), 6269--6284.

\bibitem{SHC09} 
\newblock P.~Str\'{a}nsk\'{y}, P.~Hru\v{s}ka  and P.~Cejnar,
     \newblock \emph{Quantum chaos in the nuclear collective model:
Classical-quantum correspondence},
\newblock  Phys.~Rev.~E, \textbf{79} (2009), 046202.



\end{thebibliography}
\end{document}